\documentclass[preprint,12pt]{elsarticle}
\usepackage{color}
\usepackage{amsmath}
\usepackage{amssymb}

\usepackage{graphicx}

\biboptions{sort&compress}

\journal{Computer Physics Communications}
\begin{document}
\begin{frontmatter}
\title{Optimizing Event-Driven Simulations}
\author{C. De~Michele \corref{CDM}}
\address{Dipartimento di Fisica, ''Sapienza'' Universit\`a di Roma,\\ P.le Aldo Moro 2, 00185 Roma, Italy}%
\ead{cristiano.demichele@roma1.infn.it}
\cortext[CDM]{Tel.: +390649913524; fax: +39064463158.} 



\begin{abstract}
Event-driven molecular dynamics is a valuable tool
in condensed and soft matter physics when particles can be
modeled as hard objects or more generally if their interaction potential
can be modeled in a stepwise fashion.
Hard spheres model has been indeed widely used both for computational
and theoretical description of physical systems.
Recently further developments of computational techniques allow 
simulations of hard rigid objects of generic shape.
In present paper we will present some optimizations for event-driven simulations 
that offered significant speedup over previous methods. In particular we will
describe a generalization of well known linked list method and an 
improvement on nearest neighbor lists method recently proposed by us.


\vspace{1pc}
\end{abstract}
\end{frontmatter}
\section{Introduction}

Systems composed of  many particles can be modeled as 
hard rigid bodies (HRB) if excluded volume interactions are dominant
and despite absence of any attraction they exhibit a rich phase diagram
especially if their shape is non-spherical  
\cite{DeMicheleNEMGLASS,Frenkel84}.
The spherical version of these models has already proven to be quite flexible and have been used to tackle, for example, several biological problems 
\cite{RapaportVirusPRL08,GiuEyeLensPRL07,SergeyPNAS05}. 
The generalization to non isotropic object 
will increase even more their flexibility and applicability.

Furthermore also attractive interactions between HRBs, provided that they 
are short-ranged and/or localized, can be properly modeled employing sticky spots (SS) \cite{DeMicheleSticky06, DeMicheleSilica06, DeMicheleDGEBA08}.

Several numerical techniques have been developed in the
past to perform molecular dynamics simulations of particles interacting with only excluded volume interactions. Dealing with hard bodies, the system is propagated in configuration space from one event
to next,  giving rise to so-called event-driven molecular dynamics (EDMD).
The essence of these EDMD numerical algorithms involves the evaluation of the overlap between different objects \cite{Donev05b,JiaoPRL08,Vieillard72,Perram85} or, equivalently, their geometrical distance \cite{mioJCP}.

In this paper we propose two optimizations for EDMD, namely: 
multiple linked lists (MLL) method and nearest neighbor lists with null spots  (SNL).
MLLs method proved to be very useful in simulating mixtures of hard spheres with 
very different sizes, while SNL method offers a significant speedup for simulating
particles with more complicated shapes, like hard-ellipsoids (HE)\cite{DeMicheleNEMGLASS}
or superquadrics (SQ) \cite{mioJCP,JiaoPRL08}.


\section{Simulating Hard Rigid Bodies with Sticky Spots}
\label {sec:SQsim}
HRBs can be simulated calculating their distance and collision contact point and time
using Newton-Raphson method to solve a proper set of non linear equations
as shown in \cite{mioJCP}.
In \cite{mioJCP} we have also shown how to make use of linked lists (LL)
and nearest neighbor lists (NNL) to attain good performance.
If HRBs have a pronounced non-spherical shape, i.e. a large aspect ratio,
it is mandatory to use NNL in order to minimize number of collision
predictions.

The event-calendar has been implemented using an hybrid approach as explained 
in \cite{PaulJCP2007}, where bounded priority queue is built on top of 
a binary tree implemented as in \cite{RapaBook}. All operations (insertion, deletion 
and next event retrieve) with such priority queue have a complexity $O(1)$ with respect
to particles number.
In order to avoid round-off problems for events very close to each other,
we shift forward time origin periodically.

Particles interact through an hard core potential and they may be decorated with spherical 
sticky spots (SS) that interact via a square well potential \cite{DeMicheleDGEBA08}.
Predicting next possible collision means that we have to take into account both collisions between spots and hard core collisions between HRBs. More specifically if $t_{SW}$ is the next collision time between spots of two particles $A$ and $B$ and $t_{HC}$ is the next collision time between their 
hard cores, the next event between $A$ and $B$ will occur at time
$t_{next} = \min\{t_{SW},t_{HC}\}$. 
All the details for an efficient algorithm to find collision times between spots can be found in \cite{DeMicheleSticky06}.

SSs have to be considered in calculating escape time from NNL. If $t_{HC}^{bb}$
is the escape time of the HRB from its bounding box (BB) and $t_{SS}^{bb}$ is the escape 
time of its SSs, then the escape time $t_{BB}$ of a particle from its BB will be 
$t_{BB} = \min\{t_{HC}^{bb}, t_{SS}^{bb}\}$. 
$t_{HC}^{bb}$ can be calculated evaluating the time evolution of the distance between
the HRB and its BB, i.e. $d(t) = \min_i\{d_i(t)\}$, where $i=1\ldots6$ labels 
each of the $6$ BB planes and $d_i(t)$ is the distance between the given HRB and $i$-th plane.

If the HRB is decorated with SSs we have also to consider the escape time of SSs
from its surrounding BB\footnote{Note that in this case BB must enclose both the HRB and the SSs}.  
The escape time can be calculated in this case considering simultaneous evolution of all distances
between SSs and the $6$ BB planes.

\section{A Novel Method to Compute Escape Time}
\label{sec:nullSS}
\begin{figure}
\begin{center}
\includegraphics[width=0.6\linewidth]{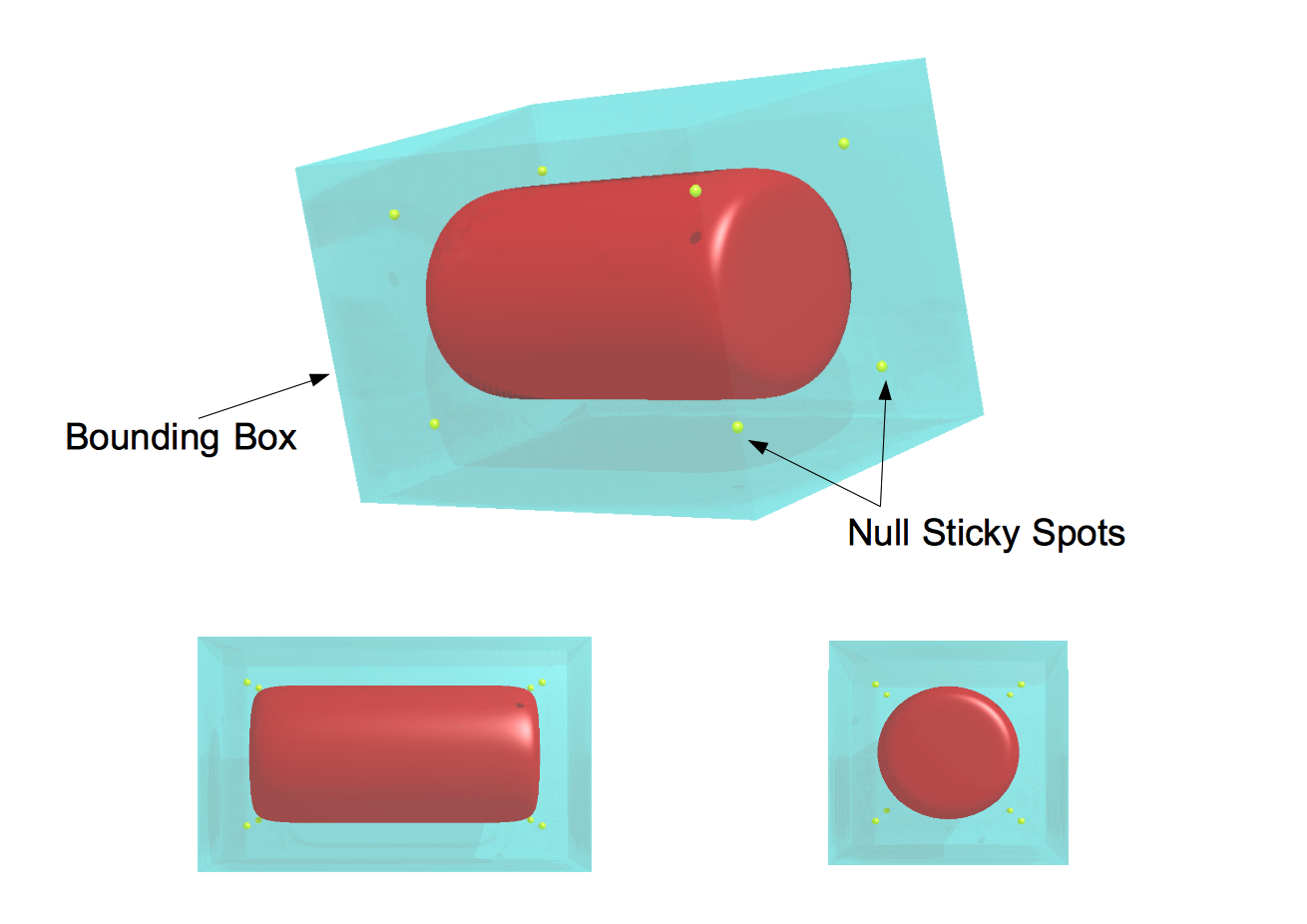}
\caption{\label{fig:nnlwithSS} SQ enclosed in its bounding box (cyan parallelepiped) 
with SS of null diameter, represented here as finite size yellow spheres. }
\end{center}
\end{figure}

In \cite{mioJCP} calculation of the escape time $t_{HC}^{bb}$ of an HRB from its enclosing BB 
requires the evaluation of collision time between the HRB and its BB.
This is computationally quite expensive although much faster than the prediction 
of the collision time between two HRBs.
A possible trick is to consider a polyhedron enclosing the HRB and to evaluate the
escape time from its BB (see Fig.\ref{fig:nnlwithSS}).
This polyhedron must be chosen to fit the HRB, i.e. given a certain shape (e.g. a parallelepiped)
it should have the smallest possible size in order to enclose completely the HRB. 
Latter requirement will ensure that the escape time of such polyhedron $t_{HC}^{sbb}$ will be an underestimate of the escape time $t_{HC}^{bb}$, i.e. $t_{HC}^{sbb} < t_{HC}^{bb}$.
Evaluation of $t_{HC}^{sbb}$ consists in calculating the smallest escape time of all vertices of the polyhedron.
At this point note that such vertices can be thought as spots of null diameter (NSS) meaning
that the same algorithm used for calculating the escape time of SSs can be employed to evaluate
the escape time $t_{HC}^{sbb}$.
In Fig. \ref{fig:nnlwithSS} the polyhedron is a parallelepiped that encloses the given cylindrical-like
HRB. If shape of particles is more complicated than the one show in Fig. \ref{fig:nnlwithSS}
simply a polyhedron with a shape that fits better the HRB shape can be used, i.e. there is no need to calculate analytically any jacobian as in the method proposed in \cite{mioJCP}.

\section{Nearest Neighbour Lists with Null Spots: Performance Results}
\label{sec:SNLperf}
In this Section we will test performance of the novel method described in Section \ref{sec:nullSS}.
We consider here superquadrics (SQ), whose surface is defined as follows:
\begin{equation}
f(x,y,z) = \left |x/a\right |^n + \left | y/b\right|^m + \left | z/c \right |^p - 1 = 0
\label{eq:SQdef}
\end{equation}
where the parameters $n,m,p$ are real numbers and $a$, $b$, $c$ are the SQ semi-axes. 
A monodisperse system of $N=512$ SQs has been simulated with $n=8$, $m=p=2$
and with two equal semi-axes, i.e. $b=c$.
Such SQs can be characterized by the elongation $X_0=a/b$ and
if elongation $X_0 < 1$ particles are called ''oblate'', while if $X_0 > 1$
particles are called ''prolate''.
For this test we have taken into account only prolate SQs, whose shape resembles that
of a cylinder with smoothed edges (see Fig. \ref{fig:nnlwithSS}).
In particular elongations $X_0=1.0, 2.0,3.0$ have been studied.
The system of prolate SQs has been simulated in a cubic box of volume $V$ with periodic boundary conditions at the volume fractions $\phi=0.20,0.30,0.35,0.40$ ($\phi=\pi X_{0}b^{3}\rho/6$, where $\rho=N/V$ is the number density).
The length of the smallest semi-axes is chosen to be the unit of lenght ($b=1.0$),
the mass of the SQ is the unit of mass ($m=1$) and the moment of inertia is spherically symmetric
and equal to $1.0$.
To create the starting configuration at a desired $\phi$, an extremely diluted crystal has been melted; afterwards, the particles have been grown independently up to the desired packing fraction (quench in $\phi$ at fixed $N$, $X_{0}$), similarly to what was done in \cite{mioJCP}.

\begin{figure}
\begin{center}
\includegraphics[width=0.49\linewidth]{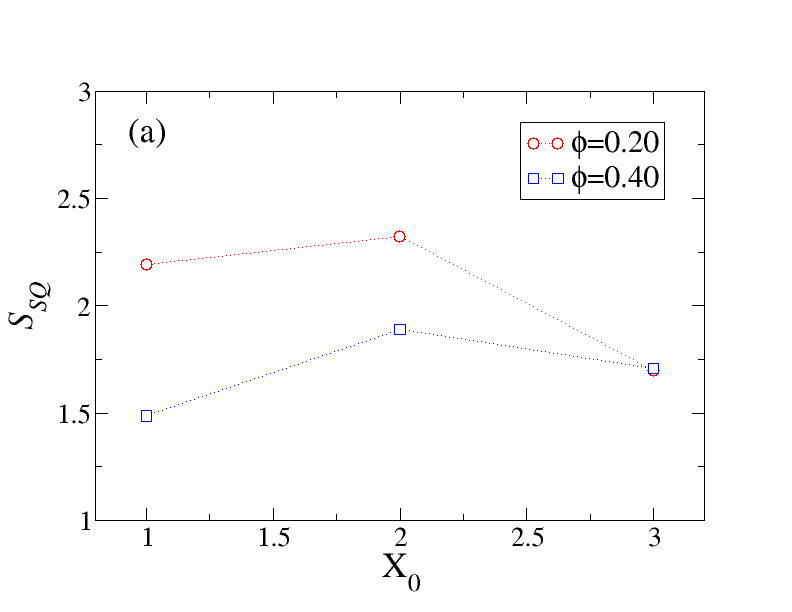}
\includegraphics[width=0.49\linewidth]{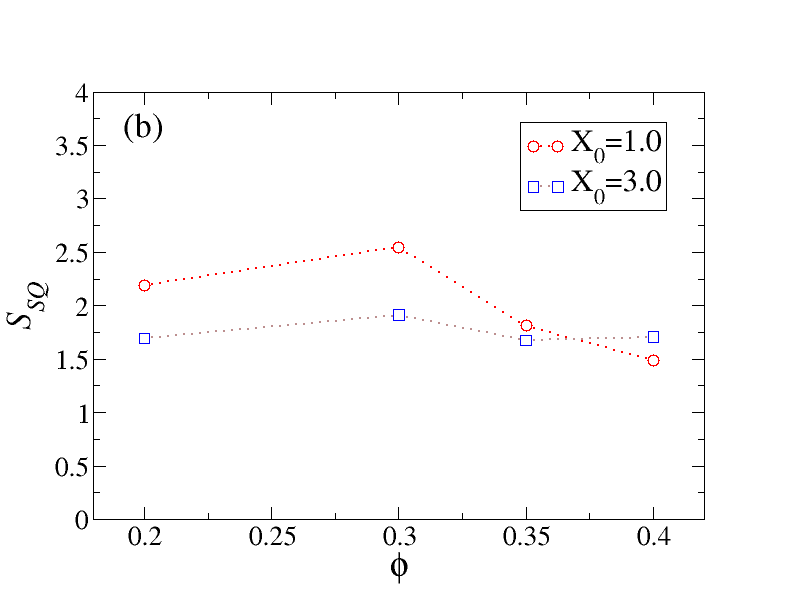}
\caption{\label{fig:SQspeedupX0_phi} (a) Speedup $S_{SQ}$ versus elongation $X_0$ 
for $\phi=0.20,0.40$ (b) Speedup  $S_{SQ}$ versus $\phi$ for $X_0=1.0, 3.0$.}
\end{center}
\end{figure}
To test the algorithm speed, we use the {\it CPU time}\footnote{{\it CPU time} means the real time spent by the  CPU for calculations.} per collision, i.e. $\tau_c = T_{tot}/N_{coll}$,
where $T_{tot}$ is the (real) time needed to perform $N_{coll}$ collisions during a simulation.

We can define the speedup $S_{SQ}$ as follows:
\begin{equation}
S_{SQ} =  \tau_c^{N} / \tau_c^{S}
\end{equation}
where $N$ refers to simulations that use NNL and $S$ refers to simulations
that make use of SNL.  
Fig. \ref{fig:SQspeedupX0_phi}  (a) shows $S_{SQ}$ as a function of elongation $X_0$ for 
two different volume fractions and it is apparent that use of SNL offers a speedup around $2$ independently of elongation.
In a similar way Fig. \ref{fig:SQspeedupX0_phi} (b) shows speedup $S_{SQ}$ for two 
elongations $X_0=1.0, 3.0$ as a function of $\phi$ and it turns out again
that $S_{SQ}$ is also independent of $\phi$.
Hence the novel method proposed for calculating the escape time of a HRB from its BB
provides a significant speedup over previous methods and this speedup is quite
independent of $X_0$ and $\phi$. This result can be fairly understood because
number of collision predictions for a given $X_0$ and $\phi$ 
are independent of method used to calculate the escape time (either SNL or NNL).
For that reason algorithm performance depends mainly on efficiency in rebuilding SNL or NNL, 
but SNL are more efficient than NNL and {\it CPU time} needed
to calculate escape time is quite insensitive to elongation and volume
fraction.


\section{Binary Mixtures of Hard-Sphers: Multiple Linked Lists}
Linked lists\footnote{they are also referred to as linked cell lists} (LL)
 \cite{RapaBook} are commonly employed in molecular dynamics simulations
in order to avoid to check all the $N^2$ possible collisions among $N$ simulated particles. 
In the LL method, the simulation box is partitioned into $M^3$ cells and only collisions 
between particles inside the same cell or its $26$ adjacent cells are accounted for. This also means that, whenever an object crosses a cell boundary going from cell $a$ to a new cell $b$, 
it has to be removed from cell $a$ and added to cell $b$.
Several variants of original LL method have been proposed in literature \cite{Germano2010}
with a view to improving the original LL method (as discussed for example in \cite{AllTildbook}).
All these improved LL methods are intended to avoid unnecessary distance
calculations but they do not tackle the case of a system composed of different species
having very different sizes.
Here we consider the case of a system composed of $N_s$ several species having different sizes
$\left \{\sigma_1, \ldots \sigma_{N_s}\right \}$ and for the sake of simplicity we can assume
that particles are spheres. 
\begin{figure}
\begin{center}
\includegraphics[width=0.65\linewidth]{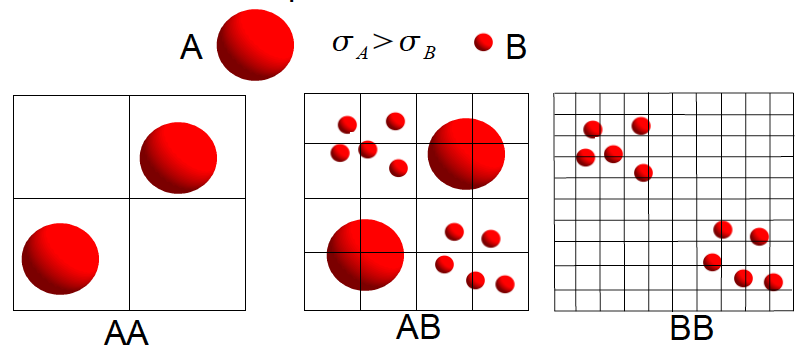}
\caption{\label{fig:MLL} Pictorial representation of Multiple Linked Lists for 
a binary mixture of two particles where diameter $\sigma_A$ of particles  $A$ is bigger
than diameter $\sigma_B$ of particles $B$. For each possible interaction $AA$, $AB$ and $BB$
a different partitioning of simulation box into cells is employed, i.e. multiple linked cell lists are used.}
\end{center}
\end{figure}

For making use of the original LL method, the simulation box must be partitioned into cubic cells and each cell must have a side length slightly greater than $\sigma_{max} = \max_i \{\sigma_i\}$.
In general, this restriction can considerably compromise algorithm performance.
To understand this, consider $m_i$ particles of diameter $\sigma_i$ inside a cell 
and  $q_i = \left (\sigma_{max}/\sigma_i \right )^3$, at a fixed volume fraction, 
if  $q_i \rightarrow 0 $ then $m_i \rightarrow \infty $ and performance are severely compromised.

A possible generalization of the original LL method for dealing with such polydisperse system consists in using one different LL for each pair of species, meaning that for each pair of species 
a different partitioning of simulation box into cells is used as illustrated pictorially in Fig. \ref{fig:MLL}
for a binary mixture of hard spheres.

In general if one has $N_s$ different species, $N_s (N_s + 1)/2$
different LLs must be used where the cell side length for interacting species $l$ and $m$
\footnote{Assuming that they are additive.} has to be greater than $(\sigma_l + \sigma_m)/2$. 
Implementation of MLL is quite straightforward except for the following remark: 
if particles, as they cross simulation box boundaries, are reinserted into
the box with periodic boundary conditions, one has to ensure
that reinsertion happens only once.
 
\section{Speedup of Multiple Linked Lists}
In this Section we will test performance of MLL method for a binary mixture of spheres having very different size. The two species (labeled by A and B) are
characterized by a diameter ratio $\sigma_A/\sigma_B = q > 1$ and their masses
are chosen to be equal and unitary. $N_A$ and $N_B$ will be number of particles
$A$ and $B$ respectively and $N=N_A+N_B$.
The number of particles $A$ will be kept fixed to $250$, i.e. $N_A=250$.
We will investigate the algorithm performance varying
$N$, the volume fraction $\phi$ and the size ratio $q$. 
The simulation box is cubic with periodic boundary conditions. As for SQs we make use of the $O(1)$ event-calendar proposed in \cite{PaulJCP2007}. 

\begin{figure}
\begin{center}
\includegraphics[width=0.49\linewidth]{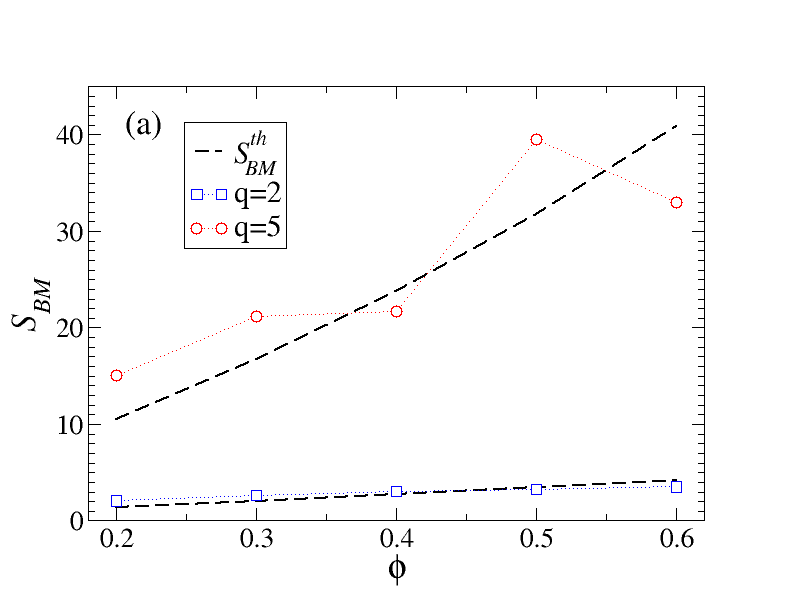}
\includegraphics[width=0.49\linewidth]{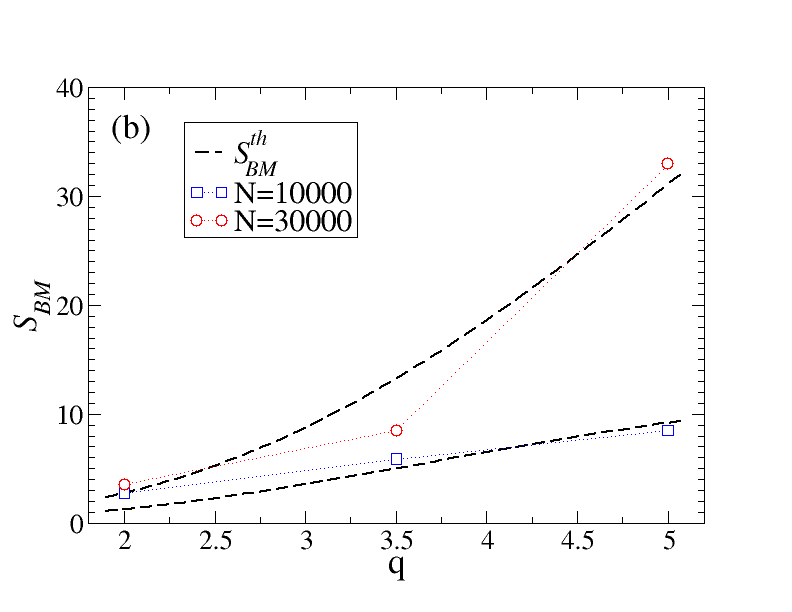}
\caption{\label{fig:BMspeedupPhi_q} (a) Speedup $S_{BM}$ versus total volume fraction 
$\phi$ for binary mixtures for $q=2, 5$ with $N=30000$ for all points.
(b) Speedup $S_{BM}$ versus size ratio $q$ for $N=10000, 30000$ with $\phi=0.60$ for all points. Dashed lines are fits to function $S^{th}_{BM}$ defined in Eq. (\ref{eq:Sth}).}
\end{center}
\end{figure}

Again we define the speedup $S_{SQ}$ as follows:
\begin{equation}
S_{BM} =  \tau_c^{LL} / \tau_c^{MLL}
\end{equation}
where $LL$ refers to simulations that use single LL, $MLL$ refers to simulations
that employ MLL and $\tau_c$ is the {\it CPU time} per collision.
Fig. \ref{fig:BMspeedupPhi_q} (a) shows $S_{BM}$  as a  function of $\phi$ for
two different size ratios $q$. It is remarkable that speedup ranges from 
$15$ to $40$.

The number of spheres of type $B$, $N_B^{cell}$, within a cell using conventional $LL$ method
is roughly:
\begin{equation}
N_B^{cell} = \frac{6\, q^3\phi }{\pi \left [q^3 (1-\phi) N_A/N_B + 1\right ]} 
\label{eq:nbcell}
\end{equation}
and a analytical estimate $S^{\it th}_{BM}$ for the speedup turns out to be \footnote{for $MLL$ method $N_B^{cell}$ is of the order of $1$.}:
\begin{equation}
\label{eq:Sth}
 S^{\it th}_{BM} = K N_B^{cell}
 \end{equation}
where $K$ is an arbitrary constant.
Fig. \ref{fig:BMspeedupPhi_q} (a) and (b) shows also fits to simulations data of function
 $S^{th}_{BM}$ defined in Eq. (\ref{eq:Sth}). Agreement between numerical data
 and $S^{th}_{BM}$ proves that $S^{th}_{BM}$ is a reasonable estimate.




\section{Conclusions}
In this paper two novel techniques easy to implement have been proposed for optimizing EDMD simulations. The SNL method offers a nearly constant speedup around $2$ with respect to the old method proposed in \cite{mioJCP} and it can be easily adapted to more complicated shapes
of simulated particles. The SNL method is actually used for simulating
a recently developed model of DNA duplexes (DNAD) \cite{BelliniScienceDNAD07} consisting in cylindrical-like SQs decorated
with two sticky spots on their two bases in order to model stacking interactions between DNADs.
The details of latter study will be given in a future publication.

MLLs have been already used in \cite{absorbPRL2010}  where spherical particles,
evolving according to event-driven brownian dynamics \cite{BD4HS}, can be absorbed by one
fixed spherical sink whose size may be much greater than diffusing particles.
MLLs are also currently used for investigating phase diagram of a binary mixture of hard spheres
whose size ratio $q=5$ upon changing the partial volume fractions of two species.
For this system it is possible to calculate a theoretical phase diagram with respect
to glass transition within the framework of Mode Coupling Theory and recently
jamming lines for such system have been evaluated both theoretically (using Replica theory) and numerically\cite{BMzampPRL2009}. In view of growing interest for this system it is crucial to have an efficient algorithm to explore whole phase diagram.
It is worth noting that also conventional time-driven molecular dynamics may benefit from
MLL.  Huge increase in performance provided by MLL, when the length scales between the components is so important, could play a relevant role in multiscale simulations, a topic that is attracting a lot of interest for biological and material applications.

Finally the two optimization methods, SNL and MLL, illustrated in this paper can be used together,
like LL and NNL, by using MLL to generate SNL.   

\section*{Acknowledgments}
The author acknowledges support from ERC (226207-PATCHYCOLLOIDS) and 
warmly thanks Prof. G. Foffi for careful reading of the manuscript.

\section*{References}
\bibliographystyle{model1-num-names}
\bibliography{optimiz-ed}

\begin{thebibliography}{21}
\expandafter\ifx\csname natexlab\endcsname\relax\def\natexlab#1{#1}\fi
\providecommand{\bibinfo}[2]{#2}
\ifx\xfnm\relax \def\xfnm[#1]{\unskip,\space#1}\fi
\bibitem[{{De~Michele} et~al.(2007){De~Michele}, Sciortino, and
  Schilling}]{DeMicheleNEMGLASS}
\bibinfo{author}{C.~{De~Michele}}, \bibinfo{author}{F.~Sciortino},
  \bibinfo{author}{R.~Schilling},
\newblock \bibinfo{title}{Dynamics of uniaxaial hard ellipsoids},
\newblock \bibinfo{journal}{Phys. Rev. Lett.} \bibinfo{volume}{98}
  (\bibinfo{year}{2007}) \bibinfo{pages}{265702}.
\bibitem[{Frenkel et~al.(1984)Frenkel, Mulder, and McTague}]{Frenkel84}
\bibinfo{author}{D.~Frenkel}, \bibinfo{author}{B.~M. Mulder},
  \bibinfo{author}{J.~P. McTague},
\newblock \bibinfo{title}{Phase diagram of a system of hard ellipsoids},
\newblock \bibinfo{journal}{Phys.\ Rev.\ Lett.} \bibinfo{volume}{52}
  (\bibinfo{year}{1984}) \bibinfo{pages}{287--290}.
\bibitem[{Rapaport(2008)}]{RapaportVirusPRL08}
\bibinfo{author}{D.~C. Rapaport},
\newblock \bibinfo{title}{Role of reversibility in viral capsid growth: A
  paradigm for self-assembly},
\newblock \bibinfo{journal}{Phys. Rev. Letters} \bibinfo{volume}{101}
  (\bibinfo{year}{2008}) \bibinfo{pages}{186101}.
\bibitem[{Stradner et~al.(2007)Stradner, Foffi, Dorsaz, Thurston, and
  Schurtenberger}]{GiuEyeLensPRL07}
\bibinfo{author}{A.~Stradner}, \bibinfo{author}{G.~Foffi},
  \bibinfo{author}{N.~Dorsaz}, \bibinfo{author}{G.~Thurston},
  \bibinfo{author}{P.~Schurtenberger},
\newblock \bibinfo{title}{New insight into cataract formation: Enhanced
  stability through mutual attraction},
\newblock \bibinfo{journal}{Phys. Rev. Letters} \bibinfo{volume}{99}
  (\bibinfo{year}{2007}) \bibinfo{pages}{198103}.
\bibitem[{Borreguero et~al.(2005)Borreguero, Urbanc, Lazo, Buldyrev, Teplow,
  and Stanley}]{SergeyPNAS05}
\bibinfo{author}{J.~M. Borreguero}, \bibinfo{author}{B.~Urbanc},
  \bibinfo{author}{N.~D. Lazo}, \bibinfo{author}{S.~V. Buldyrev},
  \bibinfo{author}{D.~B. Teplow}, \bibinfo{author}{H.~E. Stanley},
\newblock \bibinfo{title}{Folding events in the $21-30$ region of amyloid
  $\beta$-protein (a$\beta$) studied in silico},
\newblock \bibinfo{journal}{Proc. Natl. Acad. Sci.} \bibinfo{volume}{102}
  (\bibinfo{year}{2005}) \bibinfo{pages}{6015--6020}.
\bibitem[{{De~Michele} et~al.(2006{\natexlab{a}}){De~Michele}, Gabrielli,
  Tartaglia, and Sciortino}]{DeMicheleSticky06}
\bibinfo{author}{C.~{De~Michele}}, \bibinfo{author}{S.~Gabrielli},
  \bibinfo{author}{P.~Tartaglia}, \bibinfo{author}{F.~Sciortino},
\newblock \bibinfo{title}{Dynamics in the presence of attractive patchy
  interactions},
\newblock \bibinfo{journal}{J. Phys. Chem. B} \bibinfo{volume}{110}
  (\bibinfo{year}{2006}{\natexlab{a}}) \bibinfo{pages}{8064}.
\bibitem[{{De~Michele} et~al.(2006{\natexlab{b}}){De~Michele}, Tartaglia, and
  Sciortino}]{DeMicheleSilica06}
\bibinfo{author}{C.~{De~Michele}}, \bibinfo{author}{P.~Tartaglia},
  \bibinfo{author}{F.~Sciortino},
\newblock \bibinfo{title}{Slow dynamics in a primitive tetrahedral network
  model},
\newblock \bibinfo{journal}{J. Chem. Phys.} \bibinfo{volume}{125}
  (\bibinfo{year}{2006}{\natexlab{b}}) \bibinfo{pages}{204710}.
\bibitem[{Corezzi et~al.(2008)Corezzi, {De Michele}, Zaccarelli, Fioretto, and
  Sciortino}]{DeMicheleDGEBA08}
\bibinfo{author}{S.~Corezzi}, \bibinfo{author}{C.~{De Michele}},
  \bibinfo{author}{E.~Zaccarelli}, \bibinfo{author}{D.~Fioretto},
  \bibinfo{author}{F.~Sciortino},
\newblock \bibinfo{title}{A molecular dynamics study of chemical gelation in a
  patchy particle model},
\newblock \bibinfo{journal}{Soft Matter} \bibinfo{volume}{4}
  (\bibinfo{year}{2008}) \bibinfo{pages}{1173--1177}.
\bibitem[{Donev et~al.(2005)Donev, Torquato, and Stillinger}]{Donev05b}
\bibinfo{author}{A.~Donev}, \bibinfo{author}{S.~Torquato},
  \bibinfo{author}{F.~H. Stillinger},
\newblock \bibinfo{title}{Neighbor list collision-driven molecular dynamics
  simulation for nonspherical hard particles. ii. applications to ellipses and
  ellipsoids},
\newblock \bibinfo{journal}{Journal\ of\ Computational\ Physics}
  \bibinfo{volume}{202} (\bibinfo{year}{2005}) \bibinfo{pages}{765--793}.
\bibitem[{Jiao et~al.(2008)Jiao, Stillinger, and Torquato}]{JiaoPRL08}
\bibinfo{author}{Y.~Jiao}, \bibinfo{author}{F.~H. Stillinger},
  \bibinfo{author}{S.~Torquato},
\newblock \bibinfo{title}{Superdisks and the role of symmetry},
\newblock \bibinfo{journal}{PRL} \bibinfo{volume}{100} (\bibinfo{year}{2008})
  \bibinfo{pages}{245505}.
\bibitem[{Vieillard-Baron(1972)}]{Vieillard72}
\bibinfo{author}{J.~Vieillard-Baron},
\newblock \bibinfo{title}{Phase transitions of the classical hard-ellipse
  system},
\newblock \bibinfo{journal}{Journal of Chemical Physics} \bibinfo{volume}{56}
  (\bibinfo{year}{1972}) \bibinfo{pages}{4729--4744}.
\bibitem[{Perram and Wertheim(1985)}]{Perram85}
\bibinfo{author}{J.~Perram}, \bibinfo{author}{M.~Wertheim},
\newblock \bibinfo{title}{Statistical mechanics of hard ellipsoids. i. overlap
  algorithm and the contact function},
\newblock \bibinfo{journal}{J. Comp. Phys.} \bibinfo{volume}{58}
  (\bibinfo{year}{1985}) \bibinfo{pages}{409--416}.
\bibitem[{{De~Michele}(2010)}]{mioJCP}
\bibinfo{author}{C.~{De~Michele}},
\newblock \bibinfo{title}{Simulating hard rigid bodies},
\newblock \bibinfo{journal}{J. Comput. Phys.} \bibinfo{volume}{229}
  (\bibinfo{year}{2010}) \bibinfo{pages}{3276--3294}.
\bibitem[{Paul(2007)}]{PaulJCP2007}
\bibinfo{author}{G.~Paul},
\newblock \bibinfo{title}{A complexity {O(1)} priority queue for event driven
  molecular dynamics simulations},
\newblock \bibinfo{journal}{J. Comput. Phys.} \bibinfo{volume}{221}
  (\bibinfo{year}{2007}) \bibinfo{pages}{615--625}.
\bibitem[{Rapaport(2004)}]{RapaBook}
\bibinfo{author}{D.~C. Rapaport}, \bibinfo{title}{The Art of Molecular Dynamics
  Simulation}, \bibinfo{publisher}{Cambridge University Press},
  \bibinfo{year}{2004}.
\bibitem[{Welling and Germano(2010)}]{Germano2010}
\bibinfo{author}{U.~Welling}, \bibinfo{author}{G.~Germano},
\newblock \bibinfo{title}{Efficiency of linked cell algorithms},
\newblock \bibinfo{journal}{arXiv:1006.1239v1}  (\bibinfo{year}{2010})
  \bibinfo{pages}{1--13}.
\bibitem[{Allen and Tildesley(1989)}]{AllTildbook}
\bibinfo{author}{M.~P. Allen}, \bibinfo{author}{D.~J. Tildesley},
  \bibinfo{title}{Computer simulation of liquids},
  \bibinfo{publisher}{Clarendon Press}, \bibinfo{edition}{paperback 385pp}
  edition, \bibinfo{year}{1989}.
\bibitem[{Nakata et~al.(2007)Nakata, Zanchetta, Chapman, Jones, Cross, Pindak,
  Bellini, and Clark}]{BelliniScienceDNAD07}
\bibinfo{author}{M.~Nakata}, \bibinfo{author}{G.~Zanchetta},
  \bibinfo{author}{B.~D. Chapman}, \bibinfo{author}{C.~D. Jones},
  \bibinfo{author}{J.~O. Cross}, \bibinfo{author}{R.~Pindak},
  \bibinfo{author}{T.~Bellini}, \bibinfo{author}{N.~A. Clark},
\newblock \bibinfo{title}{End-to-end stacking and liquid crystal condensation
  of 6–to 20–base pair dna duplexes},
\newblock \bibinfo{journal}{Science} \bibinfo{volume}{318}
  (\bibinfo{year}{2007}) \bibinfo{pages}{1276--}.
\bibitem[{Dorsaz et~al.(2010)Dorsaz, {De~Michele}, Piazza, Rios, and
  Foffi}]{absorbPRL2010}
\bibinfo{author}{N.~Dorsaz}, \bibinfo{author}{C.~{De~Michele}},
  \bibinfo{author}{F.~Piazza}, \bibinfo{author}{P.~D.~L. Rios},
  \bibinfo{author}{G.~Foffi},
\newblock \bibinfo{title}{Diffusion-limited reactions in crowded environments},
\newblock \bibinfo{journal}{arXiv:1007.2529, Accepted by Phys. Rev. Lett.}
  (\bibinfo{year}{2010}) \bibinfo{pages}{1--4}.
\bibitem[{Scala et~al.(2007)Scala, De~Michele, and \hbox{Th.}
  Voigtmann}]{BD4HS}
\bibinfo{author}{A.~Scala}, \bibinfo{author}{C.~De~Michele},
  \bibinfo{author}{\hbox{Th.} Voigtmann},
\newblock \bibinfo{title}{Event-driven brownian dynamics for hard spheres},
\newblock \bibinfo{journal}{J.\ Chem.\ Phys.} \bibinfo{volume}{126}
  (\bibinfo{year}{2007}) \bibinfo{pages}{134109}.
\bibitem[{Biazzo et~al.(2009)Biazzo, Caltagirone, Parisi, and
  Zamponi}]{BMzampPRL2009}
\bibinfo{author}{I.~Biazzo}, \bibinfo{author}{F.~Caltagirone},
  \bibinfo{author}{G.~Parisi}, \bibinfo{author}{F.~Zamponi},
\newblock \bibinfo{title}{Theory of amorphous packings of binary mixtures of
  hard spheres},
\newblock \bibinfo{journal}{Phys. Rev. Lett.} \bibinfo{volume}{102}
  (\bibinfo{year}{2009}) \bibinfo{pages}{195701}.

\end{thebibliography}
\end{document}